\documentclass[twoside,twocolumn,10pt]{article}

\usepackage{wscg}           
\RequirePackage{ifpdf}
\ifpdf
 \RequirePackage[pdftex]{graphicx}
 \RequirePackage[pdftex]{color}
\else
 \RequirePackage[dvips,draft]{graphicx}
 \RequirePackage[dvips]{color}
\fi
\usepackage[english]{babel}     
\usepackage{booktabs}      
\usepackage{commath}
\usepackage{times}
\usepackage[modulo,switch]{lineno}
\usepackage{amssymb,amsmath,amsthm,enumitem}
\usepackage[caption=false,font=small,labelfont=sf,textfont=sf,labelformat=parens]{subfig}
\usepackage{tikz}
\usepackage{pgfplots}
\usepackage{multirow}
\usetikzlibrary{decorations.pathreplacing}
\usetikzlibrary{colorbrewer}
\usepackage{color}
\usepackage[sort&compress,numbers]{natbib}
\usepackage{flushend}
\usepackage{makecell}
\usepackage{multirow}
\usepackage{xr}
\usepackage{nopageno}       

\usepackage{xcolor,colortbl}
\usepackage[nodisplayskipstretch]{setspace}
\renewcommand{\baselinestretch}{0.963}
\title{Multi-Resolution Rendering for Computationally Expensive Lighting Effects}

\author{
\parbox{0.25\textwidth}{\centering
Simon Besenthal\\[1mm]
Ulm University\\[1mm]
\footnotesize simon.besenthal@uni-ulm.de
}
\hspace{0.05\textwidth}
\parbox{0.25\textwidth}{\centering
Sebastian Maisch\\[1mm]
Ulm University\\[1mm]
\footnotesize sebastian.maisch@uni-ulm.de
}
\hspace{0.05\textwidth}
\parbox{0.25\textwidth}{\centering
Timo Ropinski\\[1mm]
Ulm University\\[1mm]
\footnotesize timo.ropinski@uni-ulm.de
}
}

\usepackage{url}
\makeatletter
\def\Uslash{\mathbin{\mathchar`\/}\@ifnextchar{/}{\kern-.15em}{}}
\g@addto@macro\UrlSpecials{\do \/ {\Uslash}}
\def\Ucolon{\mathbin{\mathchar`:}\@ifnextchar{/}{\kern-.1em}{}}
\g@addto@macro\UrlSpecials{\do : {\Ucolon}}
\makeatother

\newcommand{\todo}[2][]{%
    \ifthenelse{\equal{#1}{}}
    {{\color{red}TODO:~#2}}
    {{\color{red}TODO~[#1]:~#2}}
}

\definecolor{color1}{RGB}{251,180,174}
\definecolor{color2}{RGB}{179,205,227}
\definecolor{color3}{RGB}{204,235,197}
\definecolor{color1dark}{RGB}{251,45,43}
\definecolor{color2dark}{RGB}{45,51,227}
\definecolor{color3dark}{RGB}{51,118,49}

\begin{document}

\twocolumn[{\csname @twocolumnfalse\endcsname

\maketitle  

\begin{abstract}
\noindent
Many lighting methods used in computer graphics such as indirect illumination can have very high computational costs and need to be approximated for real-time applications.
These costs can be reduced by means of upsampling techniques which tend to introduce artifacts and affect the visual quality of the rendered image.
This paper suggests a versatile approach for accelerating the rendering of screen space methods while maintaining the visual quality.
This is achieved by exploiting the low frequency nature of many of these illumination methods and the geometrical continuity of the scene.
First the screen space is dynamically divided into separate sub-images, then the illumination is rendered for each sub-image in an adequate resolution and finally the sub-images are put together in order to compose the final image.
Therefore we identify edges in the scene and generate masks precisely specifying which part of the image is included in which sub-image.
The masks therefore determine which part of the image is rendered in which resolution.
A step wise upsampling and merging process then allows optically soft transitions between the different resolution levels.
For this paper, the introduced multi-resolution rendering method was implemented and tested on three commonly used lighting methods.
These are screen space ambient occlusion, soft shadow mapping and screen space global illumination.
\end{abstract}

\subsection*{Keywords}
Real-Time Rendering, Multi-resolution

\vspace*{1.0\baselineskip}
}]


\section{Introduction}
\label{sec:introduction}

\copyrightspace

As a subarea of computer science, real-time computer graphics has developed continuously since the middle of the last century and is of great importance today.
With a variety of applications, including medicine or computer-aided design (CAD), real-time computer graphics is nowadays indispensable in many areas of life and is thus a relevant factor in research as well as in business.
To render a realistic image many optical and physical phenomena such as camera lenses, light transport, or micro-surface structure must be taken into account.
All of these phenomena need to be calculated at pixel level but might rely on information of the surrounding scene to create the effect.
Therefore, the number of pixels to be rendered, especially with more complex illumination, is crucial to the necessary computing power and thus to the performance of an application.
While the increase in computing power of modern graphics hardware allows for more complicated algorithms, the demand for photo-realistic global illumination effects and high output resolutions in real-time graphics can not be met by current hardware sufficiently.

In order to reduce the computational effort upsampling is often used.
This technique renders individual effects, or sometimes the full image, in a lower resolution.
Subsequently, the generated images are scaled back up to the full resolution by interpolation.
Ultimately, fewer pixels must be calculated and stored, which reduces the computational effort and also the required storage space.
Upsampling is particularly common in soft, continuous post-processing effects such as bloom filters or blur, in which quality losses are virtually invisible, depending on the scaling factor.
If, on the other hand, you render effects with more concrete structures such as shadows or reflections in a lower resolution and then scale them up, hard edges are displayed washed out and aliasing becomes visible.
In addition, there is a risk of under-sampling, which can cause visual artifacts affecting the image quality, especially in animated scenes or during camera movements.
Rendering such effects or the entire image by upsampling is therefore usually not always useful, however, two interesting observations can be made:
Although such effects may generally have more concrete structures such as hard edges, these high-frequency details are firstly not necessarily evenly distributed in the image space, and secondly, they are often only marginally present in relation to the total area.
For example, considering naive shadow mapping with a single light source, depending on the complexity of the scene, a rendered image may contain large areas that are either completely shaded or fully illuminated.
Nevertheless, the necessary operations to determine the brightness of these areas are performed for each individual pixel.
For naive shadow mapping, this is certainly not important, but if one considers computationally more complex effects such as ambient occlusion or indirect illumination, the performance could be drastically increased by an intelligent subsampling of certain image areas.

The technique developed in this work exploits the often existing optical continuity of a scene in order to realize computationally intensive lighting effects more efficiently.
For this purpose, the image space is first divided into multiple disjoint partial images, so that areas which contain edges or are in their immediate vicinity are separated from areas without edges or with a greater distance to them.
Each partial image can be rendered individually with the illumination effects to be realized in suitable resolutions.
In principle, a higher resolution is required to correctly create the effect in areas with a higher detail density.
However, areas that do not include edges and thus have a lower density of detail can be rendered in lower resolution.
The partial images are then reassembled to the original image.
In the best case, this image should not differ visually from a full-resolution rendered image.
Of particular importance for visual quality and performance is the way in which the individual steps of the technology work.
For each different step approaches are presented and explained in this paper.

\section{Related Work}
\label{sec:relatedwork}

In this section we present and explain the techniques and approaches relevant to this work.
They follow similar conceptual principles and can be considered as a starting point for the technique developed here.
We also highlight the differences to these approaches.

\subsection{Upsampling}
Upsampling is a technique commonly used in low-frequency visual effects in real-time computer graphics.
Examples of effects that are often realized are Bloom or Glare filters~\cite{Greg2004} and Depth of Field~\cite{scheuermann2004advanced}.
The blur for the respective effect is not rendered in the full resolution of the application, but in an often much lower resolution.
Subsequently, the result is scaled back to the full screen size by means of bilinear interpolation.
This can greatly increase the performance at the same optical quality.

\subsection{Adaptive Multi-Resolution}
There are several approaches that split the computation of illumination effects into multiple resolutions to separate the rendering of low frequency and higher frequency components of these effects.
Examples are implementations for indirect light transport~\cite{soler2010deferred} and Screen Space Ambient Occlusion~\cite{hoang2010multi}, which achieve better performance with optically good results.
In both approaches, multiple mipmap stages of the G-buffer are used to render the lighting effect to be realized in various resolutions.
Subsequently, an upsampling is performed by means of bilateral filters and the different levels are combined.
The multi-resolution rendering technique developed in this work makes use of the fundamental principle of separating high and low-frequency components of the illumination, but divides the image into several partial images on the basis of these different proportions.
An area of the image is not rendered in all resolutions, but in the best case only in one.
This makes it possible to drastically reduce the calculations for higher-frequency components in the image areas in which ultimately no high-frequency components occur exactly.

Nichols and Wyman~\cite{nichols2010interactive} describe a real-time technique for rendering indirect illumination using multi-resolution splatting.
They use min-max mipmaps to find the discontinuities in the geometry.
Using these discontinuities, the image space is hierarchically divided into smaller squares, so that areas with higher-frequency components obtain a finer resolution.
After the image is completely split into such `splats' of an appropriate size, the indirect illumination is rendered in all resolutions and the layers are then combined by upsampling to produce the final image.
Our technique differs from the algorithm presented by Nichols and Wyman among other things in the method used to decide which resolution to render in.
We can apply more flexible filters depending on the situation, while their approach using min-max mipmaps can only find geometric discontinuities.
We also use a different approach to combine the final images that prevents visible artifacts.
Finally, our technique is not only specialized for indirect illumination using Reflective Shadow Maps, but can also be applied and optimized for various lighting effects due to its high flexibility.

Iain Cantlay~\cite{cantlay2007gpu} describes a technique for rendering lower resolution particles offscreen and combining the result with high resolution renderings of other geometry.
In contrast to our approach, this technique can only be applied, if distinct parts of the geometry (in this case particles) are to be rendered in a fixed lower resolution while our technique is more flexible working on pixels.

Guennebaud et al.~\cite{guennebaud2007high} use variable resolutions for soft shadow mapping in screen space.
Again our approach is more flexible and can be applied to a multitude of screen space effects.

\subsection{Variable Rate Shading}
He et al.~\cite{He:2014} propose an extension of the graphics pipeline to natively support adaptive sampling techniques.
Nvidia's Maxwell and Pascal architectures have already implemented graphics hardware technologies that could speed up the rendering of an image through the use of different resolutions.
Multi-Resolution Shading~\cite{nvidiaMRS} and Lens Matched Shading~\cite{nvidiaLMS} can be applied in virtual reality applications to adapt the resolution of individual image areas to the optical properties of the physical lens that is part of the display.
For more general uses Variable Rate Shading~\cite{nvidiaVRS} (VRS) was introduced as part of the Nvidia Turing architecture.
With this technique, the image can be divided into much finer regions, which can be rendered independently in appropriate resolutions.
The regions are made up of squares with a edge length of sixteen pixels.
Possible applications include `Content Adaptive Shading' (as for example presented by Vaidyanathan et al.~\cite{Vaidyanathan:2014}), `Motion Adaptive Shading' (as for example presented by Vaidyanathan et al.~\cite{Vaidyanathan:2012}), and `Foveated Rendering' (as presented by Guenter et al.~\cite{Guenter:2012}).
In this case, the sampling rate of the image areas is selected adequately depending on the detail density, movement, or focus of the viewer.

The multi-resolution rendering technique developed in this work allows for an even finer and more flexible division of the image, since image areas do not necessarily have to consist of square tiles, but can have any desired shape.
This means that a possibly even lower part of the image must be rendered in full resolution, and the performance can be further increased.
Apart from that, in contrast to VRS, our technique allows for any number of levels and even lower sampling rates.
Our technique is also not dependent on current graphics hardware and can be implemented for widely available systems.
In our implementation we focus on the density of details in a scene (Content Adaptive Shading) to decide for the resolution to render in but we can extend our technique by using different edge detection filters or even masks that describe the geometry of lenses in virtual reality.

\subsection{Global Illumination Effects}
\label{subsec:gi-effects}
For the exemplary implementation of our technique we use three illumination effects commonly used in modern computer graphics.

Screen Space Ambient Occlusion (SSAO) is a real-time approximation of the occlusion of ambient light by local geometry.
The technique was first presented by Mittring~\cite{mittring2007finding} and further developed and improved (e.g. by Bavoil et al.~\cite{bavoil2008image}).

Shadow Mapping is an algorithm presented by Williams~\cite{williams1978casting} that allows for a fast calculation of shadow rays using a depth buffer.
Artifacts introduced by the resolution of the depth buffer can be reduced by percentage closer filtering, introduced by Reeves et al.~\cite{reeves1987rendering} that also softens the shadows edges.
A plausible penumbra can also be realized as described by Fernando~\cite{fernando2005percentage}.
The shadow map is not only sampled at a single position but at multiple neighboring locations.

Screen Space Global Illumination as, for example, described by Ritschel et al.~\cite{ritschel2009approximating} generalizes SSAO to not only dim ambient illumination but also add indirect illumination from other surfaces visible on the screen.
The light transport between chosen samples close to a pixel is calculated inducing information from the G-Buffer.

\section{Multi-Resolution Rendering}
\label{sec:multi-resolution-rendering}
Our presented multi-resolution rendering technique can be subdivided into three basic steps.
In the first step, we create a mask in screen space, based on which the image to be rendered is divided into disjoint or complementary sub-images.
In the second step, the lighting method to be implemented is rendered for each sub-image in its adequate resolution.
Finally the sub-images are combined to create the result image.
The conceptual approaches of these steps will be described in more detail below.
A visual overview of the algorithms workflow will be given in the supplementary material.

\subsection{Mask Creation}
\label{creation-of-the-edge-image}
\begin{figure}[tb]
    \centering
    \includegraphics[width=\linewidth]{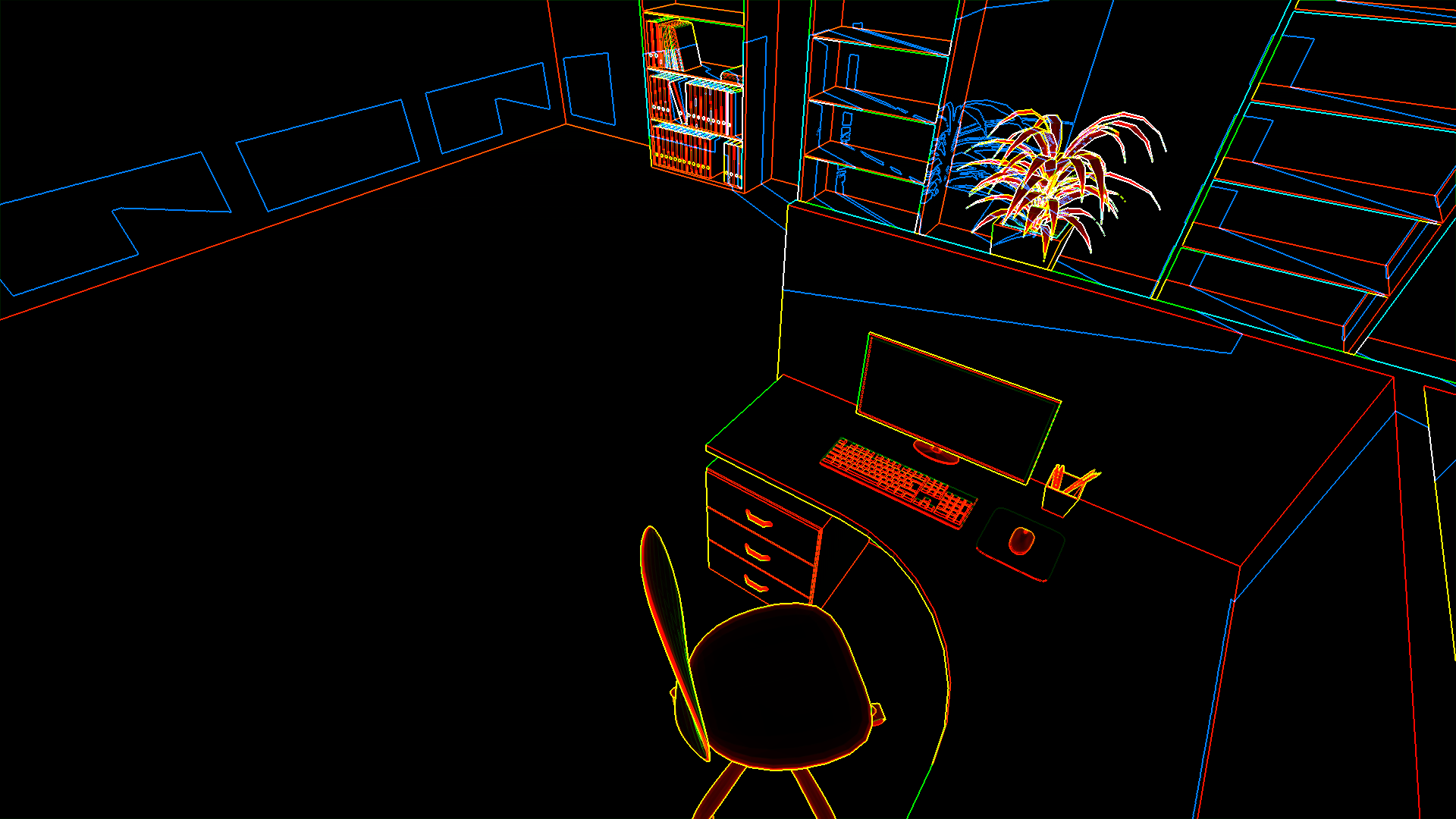}
    \caption{A possible edge image for the multi-resolution rendering technique, the edges are colored for better visualization: the red edges were determined by the differentiation of the normals, the green ones by the depth values and the blue ones by the shadows, normal edges and depth edges are often determined at the same point in the image space (yellow edges).}
    \label{fig:edges}
\end{figure}
The masks are used to divide an image into individual sub-images.
While masks can be acquired in multiple ways and even combined using the minimum or maximum (depending on the application) an obvious choice is to use them to separate the higher-frequency image parts from the low-frequency ones.
It is often sufficient to use the geometry edges of the scene in screen space to achieve this.
These can be found through the information available in the G-Buffer by numerically differentiating depth values and normals for each pixel.
For the normal, the first derivative in each of the two dimensions is sufficient, whereas for the depth values, the second derivative gives more reliable results.
The discontinuities found reproduce the geometric edges of the scene and can be used to split the image.
For screen space ambient occlusion and screen space global illumination, the geometric edges are already sufficient but depending on the illumination effect to be realized, additional information may be required.
In case of soft shadow mapping for example, the shadow edges of the scene are needed above all.
To this purpose, when creating the mask using the previously created shadow map, a fast shadow calculation (one sample per pixel) can be implemented.
We differentiate these values to find discontinuities in the shading.
To avoid artifacts at the geometry edges, we also take them into account for the mask when rendering the soft shadows.
Fig.~\ref{fig:edges} shows an edge image of a scene in which normals, depths, and shadows are differentiated.
As an alternative to the edge images we use, min-max mipmaps can also be used to decompose the image as explained by Nichols and Wyman~\cite{nichols2010interactive}.

After we created the final high-resolution mask we downsample it to the resolutions we want our final sub-images to be.
We use blur filters with different variances ($\sigma^2$) on the downsampled images to determine the areas near the edges.
The blurs variance gives the developer control over the size of the area around the edges and determines which areas around the edges are rendered in which resolution.
The variances we use can be found in Tab.~\ref{tab:weightsvariances}.

\begin{figure}[htb]
    \centering
    \includegraphics[width=\linewidth]{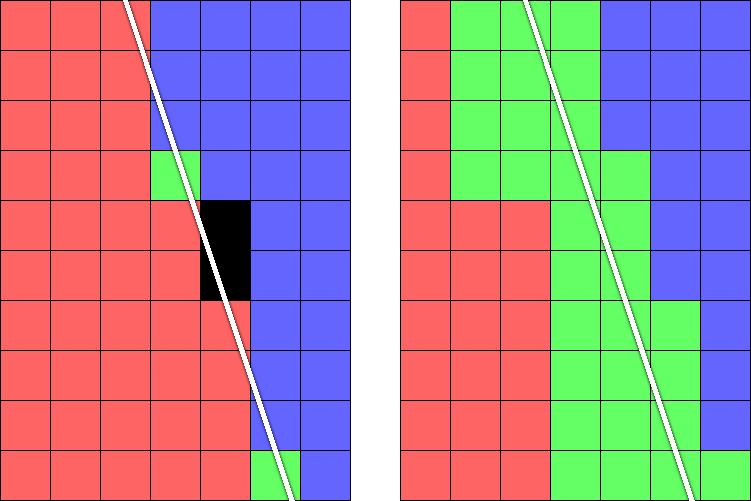}
    \caption{Without accounting for overlap (left), \glqq dead pixels\grqq\ (black) occur at the edges of the sub-images (red and blue), which are not contained in any of the sub-images and thus are not rendered.
    When ensuring an overlap (right), the intersection of the sub-images (green) prevents this circumstance.}
    \label{fig:edge-overlap}
\end{figure}

\begin{table}
    \centering
    \begin{tabular}{c||c|c||c|c||c|c}
        $i$ & \multicolumn{2}{c||}{SSAO} & \multicolumn{2}{c||}{SSM} & \multicolumn{2}{c}{SSGI} \\ 
        \hline 
        & $\sigma_i^2$ & $w_i$ & $\sigma_i^2$ & $w_i$ & $\sigma_i^2$ & $w_i$ \\ 
        \hline\hline
        1 & $0.924$ & $100$ & $0.924$ & $1000$ & $0.924$ & $1000$ \\ 
        \hline 
        2 & $1.848$ & $50$ & $1.848$ & $1000$ & -- & -- \\ 
        \hline 
        3 & $3.696$ & $20$ & $3.696$ & $1000$ & $0.924$ & $100$ \\ 
        \hline 
        4 & $0$ & $1$ & $0$ & $1$ & $0$ & $1$ \\ 
    \end{tabular} 
    \caption{Variances ($\sigma_i^2$) and weights ($w_i$) for each sub-image ($i$) of all techniques we used.
        The variances are used to blur the mask, while the weights are used to combine the final image.
        For SSGI we did not use the second sub-image at all.
    }
    \label{tab:weightsvariances}
\end{table}

\begin{figure}[htb]
  \centering
  \includegraphics[width=\linewidth]{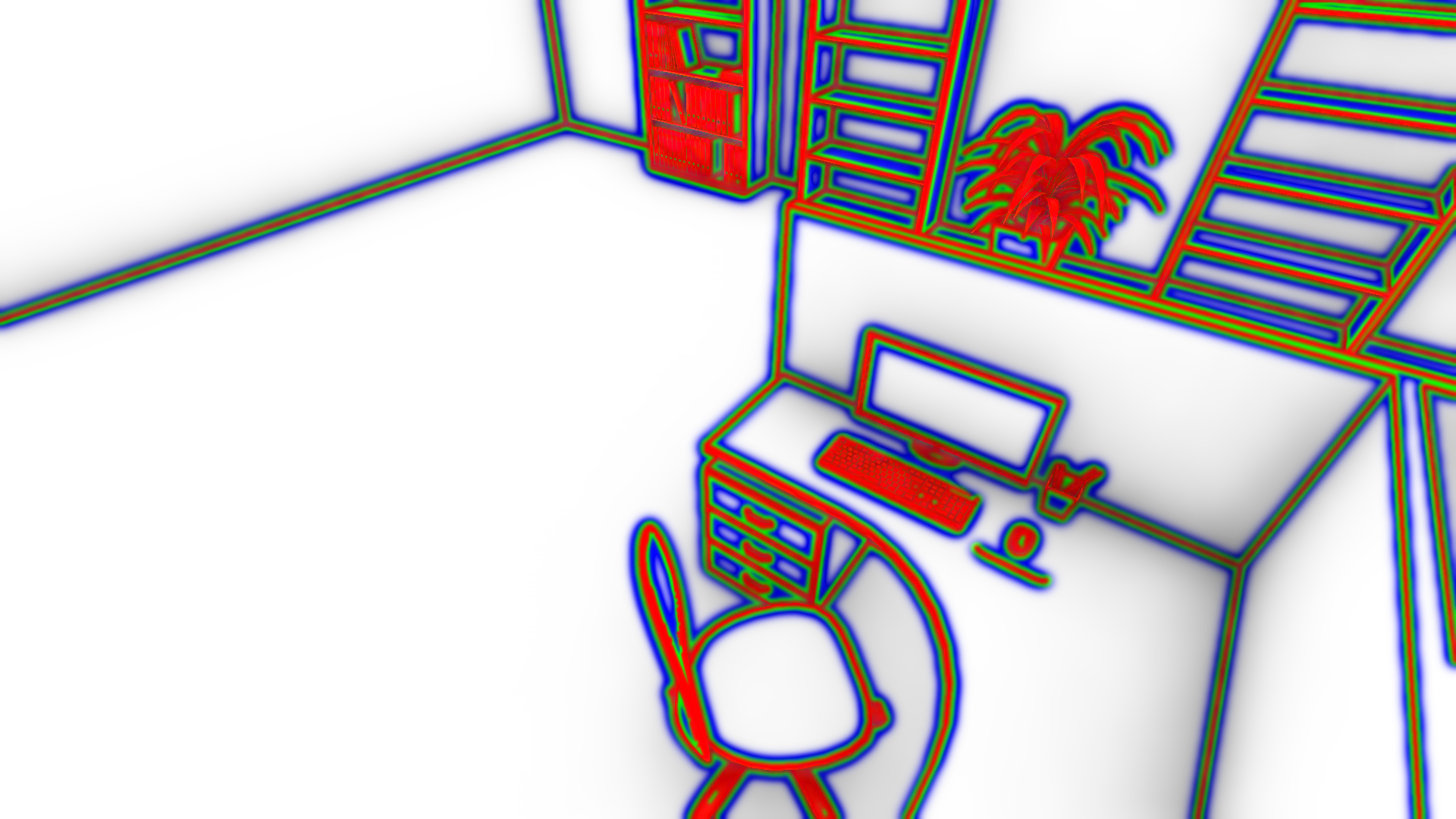}
  \caption{Visualization of the decomposition of an image into four sub-images by means of inclusive areas: The sub-image of the full resolution contains all the red areas, the sub-image of the half resolution all red and green areas, the sub-image of the quarter resolution all red, green and blue areas. The fourth sub-image renders the entire image space at an eighth of the resolution.}
  \label{fig:separate-b}
\end{figure}

A simple way to separate the image into sub-images is to divide them into complementary tiles.
An advantage of this method is the disjoint decomposition, whereby no area of the image has to be rendered multiple times.
A drawback, however, is that the granularity of the decomposition of the image is limited by the lowest resolution of a sub-image.
When naively using the granularity that is determined directly by the resolution of each sub-image, we obtained undefined spaces in the final image between two masked areas.
To avoid these we make sure areas of different resolutions have an overlap as shown in Fig.~\ref{fig:edge-overlap}.
Therefore, we do not separate the image into almost disjoint areas, but always completely include the higher resolution levels in the underlying ones.
This means, in particular, that the lowest resolution sub-image always renders the effect to be realized for the entire image.
Losses in performance due to the multiple rendering of some image areas are extremely small, because the additional computational effort arises mainly in the lower resolutions.
If the blur is optimally selected for the creation of the masks, this approach lets us keep the areas of the higher resolution levels extremely small, resulting in an overall good performance.
In addition, this decomposition approach later allows for a very simple re-composition of the final image, because the masks together with fixed weights can serve as an alpha channel for blending the sub-images (see Section~\ref{ch:blending-the-subimages}).
Fig.~\ref{fig:separate-b} shows a possible decomposition of an example scene in screen space.

\subsection{Rendering the Sub-images}
\begin{figure}[htb]
  \centering
  \includegraphics[width=\linewidth]{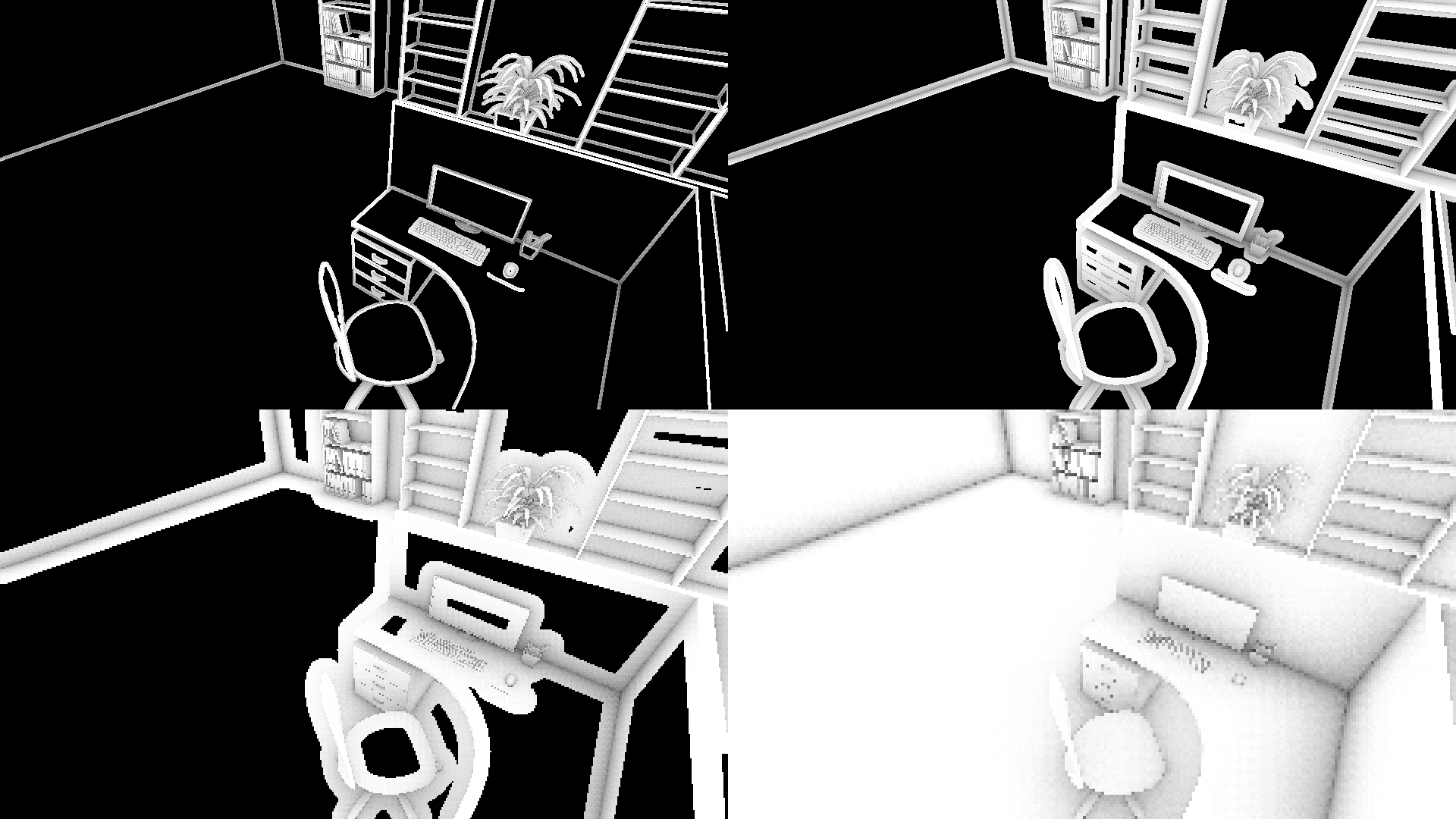}
  \caption{Screen space ambient occlusion rendered in four sub-images, no lighting is calculated for the black areas. The individual sub-images render SSAO in full (top left), half (top right), quarter (bottom left) and eighth resolution (bottom right).}
  \label{fig:stencil-render}
\end{figure}

Throughout the rendering process we generate all sub-images independently of each other in the chosen resolution.
Shape and resolution of the sub-image are defined by the masks determined in step one.
Accordingly, an image area of a sub-image is only rendered if and only if the corresponding mask in this image area permits it.
Fig.~\ref{fig:stencil-render} shows an example of rendering four sub-images.

\subsection{Blending the Sub-Images}
\label{ch:blending-the-subimages}
As the final step of the technique we blend the individually rendered sub-images in order to generate the final image.
All sub-images are upsampled to the full resolution and combined.
Using a simple bilinear interpolation would lead to artifacts, as pixels containing visual information can be interpolated with those that contain no information.

A simple solution for this problem would be bilateral interpolation as described by Tomasi and Manduchi~\cite{tomasi1998bilateral}.
When using this, the sub-images are gradually scaled and merged without scattering missing information of a resolution level into the relevant pixels of the image.
To this purpose, a sub-image is always combined with the sub-images already blended in one step.
This upsampling technique is also used by Nichols and Wyman \cite{nichols2010interactive}.

In our case we can use the decomposition masks to calculate the final blending weights.
Each sub-image, starting at the lowest resolution, is blended with the next higher resolution sub-image based on the alpha value of each mask.
The softness of the transitions between the resolution levels can be determined flexibly using weights.
These weights are multiplied with the alpha mask and define the final alpha value for blending.

\section{Implementation}
\label{sec:implementation}
In our implementation we applied our multi-resolution rendering technique to three illumination effects commonly found in modern real-time computer graphics.
These effects are SSAO, soft shadow mapping (SSM) and screen space global illumination (SSGI).
In this section, we describe the implementation of our technique and specific adjustments for the illumination effects used.
Our implementation relies solely on the OpenGL 3.3 core profile and can as such run on widely available hardware.
According to our experiences during the development stage, a decomposition in four sub-images appears as the best compromise between image quality and speed.
The width of the sub-images is successively halved, starting at full resolution width, and are set to full, half, quarter, and eighth.
For SSGI we found that not using the halved sub-image did not result in worse image quality. This contributed to a further performance enhancement.

\subsection{Rendering of the Sub-Images}
To render the sub-images, we use the previously generated masks to create a stencil buffer for each resolution determining the areas.
We check if the mask is greater than zero and set the stencil value to one or zero accordingly.
We thought about using different thresholds for creating the stencil masks but for our purposes just using zero provided the best results.
For each resolution level used, we subsequently render each sub-image using the stencil buffer to eliminate regions that we do not want to render.

For SSAO, depending on the number of samples used, we blur the resulting sub-images in order to reduce the occurring variance of the effect, especially in the lower resolutions.
However, we needed to ensure not to transport missing pixel information into the defined areas of the respective sub-image.
We achieved this, with a bilateral blur filter.

\subsection{Blending of the Sub-Images}
Subsequently, the rendered sub-images are blended to compose the final image.
We use bilinear interpolation to scale the sub-images to full size and then combine them sequentially, starting at the lowest resolution level.
We carry out the final blending between two sub-images by using the values of our masks ($a_i$) multiplied by a weight ($w_i$) as a linear interpolation parameter.
The weights of our example cases can be found in Tab.~\ref{tab:weightsvariances}.
We calculate the following for each pixel of the final image.
We define $c_i$ as that pixels color value in the $i$-th sub-image, where $c_1$ is the full resolution image.
The composed image including the $i$-th sub-image as its highest resolution is called $c'_i$.
The fourth sub-image has the lowest resolution, covers the entire image space and is defined for each pixel.
We use its value as the initial value $c'_4=c_4$.
All other $c'_i$ are calculated successively using the alpha values $a_i$ from the corresponding masks and the weights $w_i$ by:

\begin{equation}
c'_i=c_i\cdot\min(a_i w_i, 1) + c'_{i-1}\cdot\left(1-\min(a_i w_i, 1)\right)
\end{equation}

The last computed value $c_1'$ describes the pixel value of the final composite image.

\section{Evaluation}
\label{sec:evaluation}
For a basic evaluation we applied our multi-resolution rendering technique to the three illumination effects mentioned (SSAO, SSM and SSGI).
We used three test scenes ``Office'' (20,189 triangles), ``Hall'' (183,333 triangles), and ``Breakfast Room'' (a slightly modified version of the one provided by Morgan McGuire~\cite{McGuire2017Data} with 269,565 triangles) with eight camera configurations for speed and visual comparison.
For Soft Shadow Mapping and Screen Space Global Illumination, a modified version of the second scene with 255,432 triangles was used, because it works better with the given directional light sources.
For each perspective, the rendering speed was measured using our technique and compared to the speed measured for naive rendering in full resolution.
In addition, comparison images of the test scenes are shown and their differences measured and visualized.
All tests were performed on a Nvidia Geforce GTX 1080.

\subsection{Rendering Speed}
\label{subsec:speed}
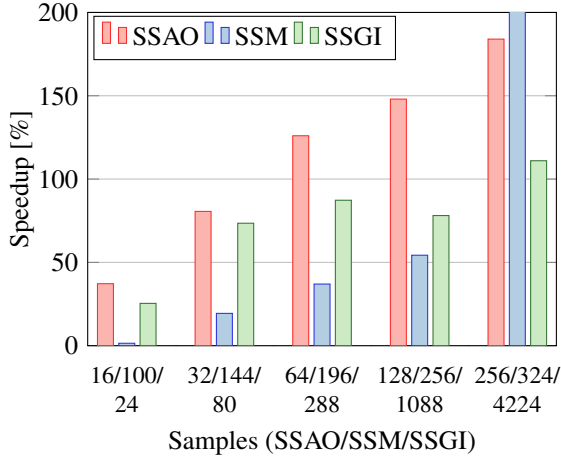
\begin{figure}
    \begin{tikzpicture}
    \begin{axis}[ybar,
    width=1.0\linewidth,
    bar width=0.21cm,
    height=6cm,
    ymin=0,
    ymax=200,
    ylabel style={yshift=-2.3ex},ylabel={Speedup [$\%$]},
    xlabel style={yshift=-3ex},xlabel={Samples (SSAO/SSM/SSGI)},
    ymajorgrids=true,
    xtick=data,
    xticklabels = {
        16/100/\\24,
        32/144/\\80,
        64/196/\\288,
        128/256/\\1088,
        256/324/\\4224
    },
    xticklabel style={yshift=-0ex,text width=1.4cm,align=center,font=\small},
    major x tick style = {opacity=0},
    minor x tick num = 1,
    minor tick length=0ex,
    legend style={at={(0.335,0.99)},
        anchor=north,
        legend columns=-1}]
    \addplot[color1dark, fill=color1] coordinates {(0,37.2) (1,80.6) (2,126) (3,148) (4,184)};
    \addplot[color2dark, fill=color2] coordinates {(0,1.46) (1,19.4) (2,37.0) (3,54.3) (4,368)};
    \addplot[color3dark, fill=color3] coordinates {(0,25.4) (1,73.5) (2,87.3) (3,78.1) (4,111)};
    \legend{SSAO, SSM, SSGI} \end{axis}
    \end{tikzpicture}
    \caption{Average speedup in percent by using our multi resolution technique in 4K (3840x2160 Pixels).
        We show the speedup for our three tested techniques using different numbers of samples for each of them.}
    \label{fig:performance-4k}
\end{figure}

For testing the speedup of our technique we used $3840\times2160$ as a base resolution.
We tested each technique with a different number of samples.
The average results for 24 different configurations (scene and camera) are listed in Fig.~\ref{fig:performance-4k}.
Despite the additional rendering steps needed, our technique outperforms naive rendering in all cases.
For a higher number of samples our technique will perform better, since more processing on the GPU can be skipped due to lower resolution rendering.

\begin{figure}
    \begin{tikzpicture}
    \begin{axis}[ybar,
    width=1.0\linewidth,
    bar width=0.21cm,
    height=6cm,
    ymin=-20,
    ymax=150,
    ylabel style={yshift=-2.3ex},ylabel={Speedup [$\%$]},
    xlabel style={yshift=-3ex},xlabel={Resolution},
    ymajorgrids=true,
    xtick=data,
    xticklabels = {
        $1280\times720$,
        $1920\times1080$,
        $2560\times1440$,
        $3840\times2160$
    },
    xticklabel style={yshift=-0ex,text width=1.4cm,align=center,font=\small},
    major x tick style = {opacity=0},
    minor x tick num = 1,
    minor tick length=0ex,
    legend style={at={(0.335,0.99)},
        anchor=north,
        legend columns=-1}]
    \addplot[color1dark, fill=color1] coordinates {(0,28.7) (1,61.1) (2,89.4) (3,126)};
    \addplot[color2dark, fill=color2] coordinates {(0,-17.4) (1,-0.621) (2,17.3) (3,37.0)};
    \addplot[color3dark, fill=color3] coordinates {(0,27.9) (1,53.4) (2,87.3) (3,138)};
    \legend{SSAO, SSM, SSGI} \end{axis}
    \end{tikzpicture}
    \caption{Average speedup in percent of our multi resolution technique at different resolutions. We used fixed numbers of samples for all techniques: 64 samples for SSAO, 196 samples for SSM, and 228 samples for SSGI.}
    \label{fig:performance-res}
\end{figure}
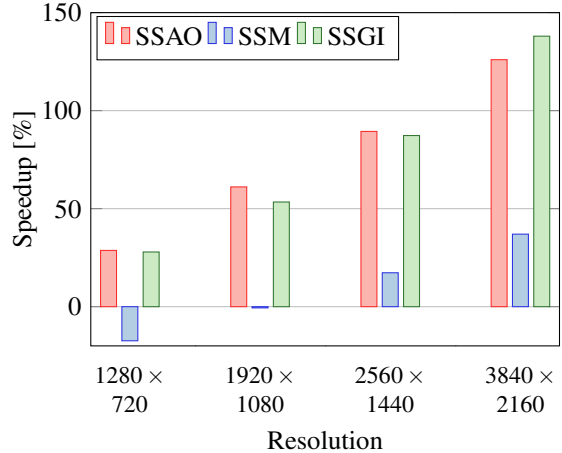

We also tested our technique for lower resolutions. The Results were not as good as the ones reported for 4K.
Nevertheless with the exception of SSM with $196$ Samples we achieved clear positive speedups for all illumination techniques even in 720p. Starting from 1440p, all illumination techniques provided positive speedups.
Our results for SSM can be explained by the fact that the technique is relatively simple while the mask generation still produces observable overhead.
Compared to this overhead, the reduction in GPU computations is relatively low.
For lower resolutions the overhead of generating the mask to divide the image and the cost of the additional rendering passes for multiple resolutions dominate over the positive effect of our technique.
Fig.~\ref{fig:performance-res} shows these results.

\subsection{Visual Comparison}
\label{subsec:visual}
\begin{figure}[b]
    \centering
    \includegraphics[width=\linewidth]{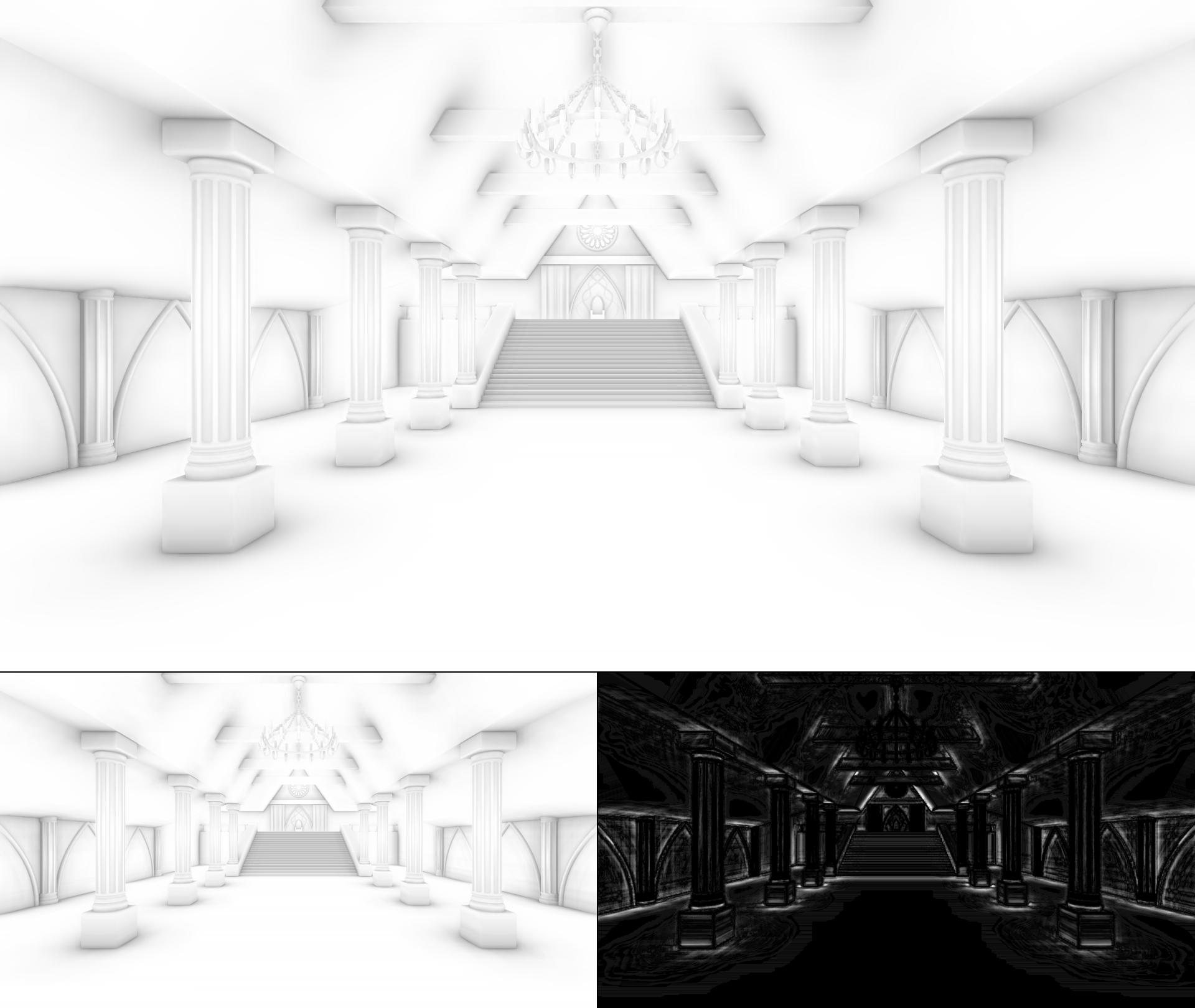}
    \caption{The ``Hall'' dataset using SSAO and 64 samples. The top image shows our multi resolution technique while in the lower left corner the reference image is shown. In the lower right corner is an enhanced difference image between those two.}
    \label{fig:ao-compare}
\end{figure}
\begin{figure}
    \centering
    \includegraphics[width=\linewidth]{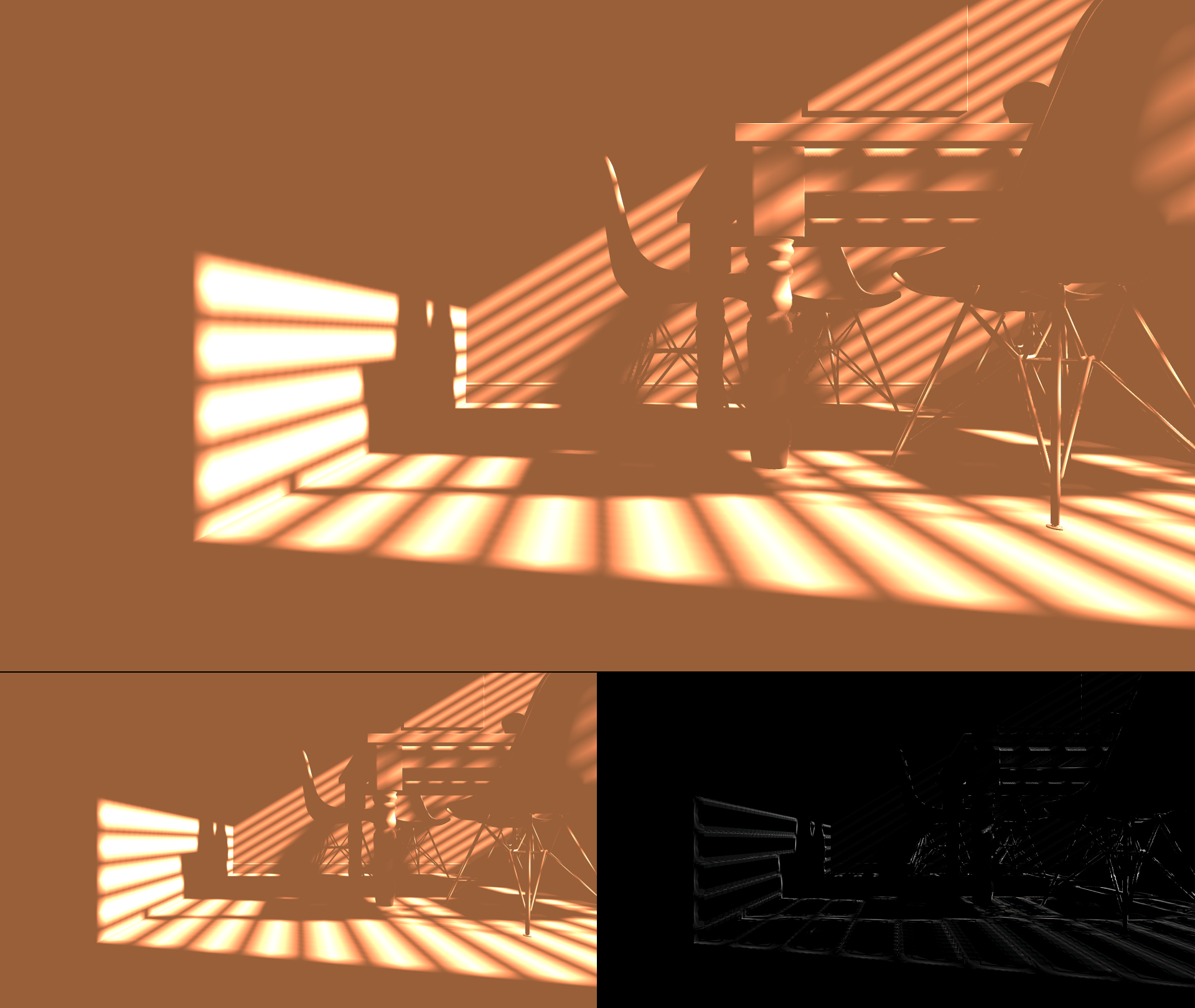}
    \caption{The ``Breakfast Room'' dataset using SSM and 196 samples. The top image shows our multi resolution technique while in the lower corner the reference image is shown. In the lower right corner is an enhanced difference image between those two.}
    \label{fig:ssm-compare}
\end{figure}
\begin{figure}
    \centering
    \includegraphics[width=\linewidth]{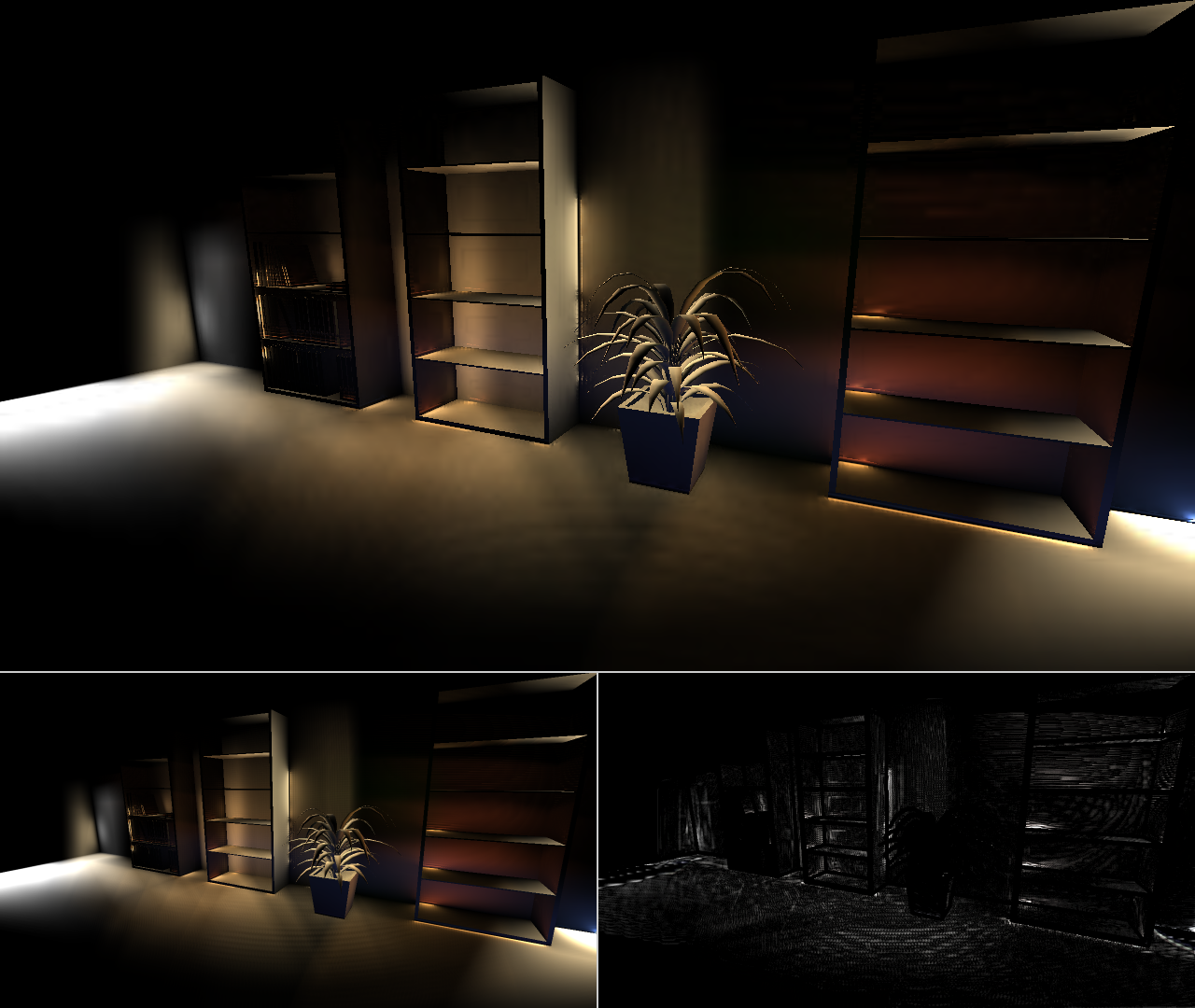}
    \caption{The ``Office'' scene using SSGI and 4224 samples. The top image shows our multi resolution technique while in the lower corner the reference image is shown. In the lower right corner is an enhanced difference image between those two. The image only shows the SSGI effect without direct illumination to better show the differences caused by our technique.}
    \label{fig:ssgi-compare}
\end{figure}

While our technique tries to prevent producing images that differ from renderings created with naive full resolution rendering, we could not prevent all visual artifacts.
As can be seen in Fig.~\ref{fig:ao-compare} to \ref{fig:ssgi-compare} these errors occur at the borders of our masks and are mostly due to the Gaussian blur we need to apply to the images to reduce discontinuities at these edges.
The blur kernel is very narrow so it is hard to detect the errors when just comparing the images directly but is visible in the difference images provided.

Fig.~\ref{fig:ao-compare} shows the results for SSAO using 64 samples.
We chose this number of samples as we think it is a reasonable choice for real applications and a good compromise between speed and image quality.
As this image is very bright the differences in the difference image are also more prominent as with the other technique.

Fig.~\ref{fig:ssm-compare} shows the results for SSM using 196 samples.
For this lighting effect we can use masks that do not depend directly on the screen space geometry for our technique.
The occurring errors are relatively low compared to the other techniques due to the parts of the scene in shadow that are lit with a constant ambient illumination.

Results of the SSGI technique we implemented are shown in Fig.~\ref{fig:ssgi-compare}.
For a visually plausible global illumination effect in screen space we needed a lot of samples so we chose to present the results for 4224 samples.
While our results are still convincing some small artifacts can be seen in the corners of the right rack.
While these present visible differences to the original image the effects are very minor.

\begin{figure}[b]
    \begin{tikzpicture}
    \begin{axis}[ybar,
    width=1.0\linewidth,
    bar width=0.21cm,
    height=6cm,
    ymin=0,
    ymax=0.02,
    ylabel style={yshift=-2.3ex},ylabel={RMS (Absolute)},
    xlabel style={yshift=-3ex},xlabel={Resolution},
    ymajorgrids=true,
    xtick=data,
    xticklabels = {
        $1280\times720$,
        $1920\times1080$,
        $2560\times1440$,
        $3840\times2160$
    },
    xticklabel style={yshift=-0ex,text width=1.4cm,align=center,font=\small},
    major x tick style = {opacity=0},
    minor x tick num = 1,
    minor tick length=0ex,
    legend style={at={(0.335,0.99)},
        anchor=north,
        legend columns=-1}]
    \addplot[color1dark, fill=color1] coordinates {(0,0.00829872) (1,0.00731286) (2,0.0066366) (3,0.00596988)};
    \addplot[color2dark, fill=color2] coordinates {(0,0.0144763) (1,0.0130612) (2,0.0123145) (3,0.010778)};
    \addplot[color3dark, fill=color3] coordinates {(0,0.00762902) (1,0.00735692) (2,0.00857135) (3,0.0111892)};
    \legend{SSAO, SSM, SSGI} \end{axis}
    \end{tikzpicture}
    \caption{The absolute root mean squared (RMS) errors between result images of our multi resolution technique and images naively rendered with high resolution.
        We used 64 samples for the SSAO images, 196 samples for SSM and 288 samples for the SSGI images.
        Values in the compared images ranged from $0$ to $1$ so the resulting errors can be considered low.}
    \label{fig:error}
\end{figure}
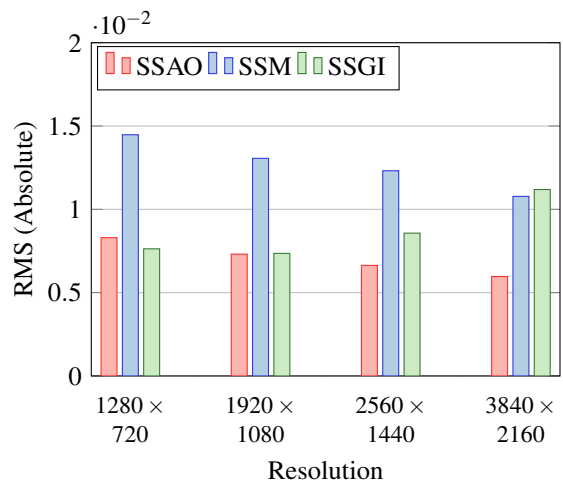

Besides the visual results we provide an overview over all errors in the graphs in Fig.~\ref{fig:error}.
These numbers do not only include the presented images but include images from all three scenes with eight camera configurations each.
These numbers support our claim that the errors introduced by our technique are very low.

\subsection{Discussion}
\label{subsec:discussion}
We presented the performance and visual quality of our method and have two general findings.
As a general rule, it was observed that illumination techniques that are more computationally demanding can benefit more from our technique than less demanding ones.
This is because of a constant overhead due to mask generation and multiple rendering passes.
This overhead becomes dominant for techniques that are less computationally demanding.
The second finding is the fact that our technique excels especially in higher resolutions for the same reason.

A minor finding is that masks which are more complicated to generate than by simply using the G-Buffer also cause a greater overhead.
This makes the use of these masks only feasible for the highest resolutions or techniques that are more computationally demanding than the soft shadow mapping presented here.

\section{Conclusion \& Future Work}
\label{sec:conclusion}
We presented a technique for multi resolution rendering that can be implemented on widely available graphics hardware.
Our technique can improve the rendering speed of screen space algorithms drastically (especially for high resolutions) as we have shown for three cases.
While the technique presented here is only used for `Content Adaptive Shading' we can trivially extend it to `Foveated Rendering' by modulating the mask we use by an importance mask provided by eye trackers.
Including `Motion Adaptive Shading' is also possible by using information of pixel motion in the mask generation process.

To further improve our technique we think that the mask generation process should be modified.
For determining the geometry edges, we use normals and depth values from the G-Buffer in screen space.
In practice however, non-smooth, modified normals are often used to calculate the illumination.
For smooth shading, pixel normals are calculated by the linear interpolation of vertex normals, but in real applications bumpmaps or normal maps are used to modify the normals.
In this case, the edge filter could potentially find many more edges, which can result in dramatically increased computational effort and significantly lower efficiency.
Possible solutions to these problems would be the exclusive use of unmodified normals or an alternative determination of the edges using the pixel locations in world space.
Another problem may arise with certain effects, including, for example, reflections or caustics, since their edges can not be calculated with the information contained in the G-buffer.
Also in this case, image areas with higher-frequency components could be rendered in too low a resolution.
For such lighting effects, further development of the progressive decomposition of the image would certainly be beneficial.
To prevent sub-sampling for some effects, sub-images could also be realized by just using a lower number of samples in full resolution instead of rendering the effect in a lower resolution.

Another interesting application for the multi resolution rendering technique would be using ray tracing for physically correct illumination.
In particular, diffuse indirect illumination can only be achieved by relatively high computational effort and can barely be realized in real-time on current graphics hardware.
Using the multi-resolution approach, the performance could be increased drastically.


\renewcommand{\baselinestretch}{1.0}
\bibliographystyle{IEEEtranN}
{\footnotesize\bibliography{multiresolution}}

\begin{thebibliography}{22}
\providecommand{\natexlab}[1]{#1}
\providecommand{\url}[1]{#1}
\csname url@samestyle\endcsname
\providecommand{\newblock}{\relax}
\providecommand{\bibinfo}[2]{#2}
\providecommand{\BIBentrySTDinterwordspacing}{\spaceskip=0pt\relax}
\providecommand{\BIBentryALTinterwordstretchfactor}{4}
\providecommand{\BIBentryALTinterwordspacing}{\spaceskip=\fontdimen2\font plus
\BIBentryALTinterwordstretchfactor\fontdimen3\font minus
  \fontdimen4\font\relax}
\providecommand{\BIBforeignlanguage}[2]{{%
\expandafter\ifx\csname l@#1\endcsname\relax
\typeout{** WARNING: IEEEtranN.bst: No hyphenation pattern has been}%
\typeout{** loaded for the language `#1'. Using the pattern for}%
\typeout{** the default language instead.}%
\else
\language=\csname l@#1\endcsname
\fi
#2}}
\providecommand{\BIBdecl}{\relax}
\BIBdecl

\bibitem[James(2004)]{Greg2004}
G.~James, \emph{{GPU Gems}}.\hskip 1em plus 0.5em minus 0.4em\relax Pearson
  Higher Education, 2004, ch. {Real-Time Glow}.

\bibitem[Scheuermann et~al.(2004)]{scheuermann2004advanced}
T.~Scheuermann \emph{et~al.}, ``{Advanced Depth of Field},'' \emph{GDC 2004},
  vol.~8, 2004.

\bibitem[Soler et~al.(2010)Soler, Hoel, and Rochet]{soler2010deferred}
C.~Soler, O.~Hoel, and F.~Rochet, ``{A Deferred Shading Pipeline for Real-Time
  Indirect Illumination},'' in \emph{ACM SIGGRAPH 2010 Talks}.\hskip 1em plus
  0.5em minus 0.4em\relax ACM, 2010, p.~18.

\bibitem[Hoang and Low(2010)]{hoang2010multi}
T.-D. Hoang and K.-L. Low, ``{Multi-Resolution Screen-Space Ambient
  Occlusion},'' in \emph{Proceedings of the 17th ACM Symposium on Virtual
  Reality Software and Technology}.\hskip 1em plus 0.5em minus 0.4em\relax ACM,
  2010, pp. 101--102.

\bibitem[Nichols and Wyman(2010)]{nichols2010interactive}
G.~Nichols and C.~Wyman, ``{Interactive Indirect Illumination Using Adaptive
  Multiresolution Splatting},'' \emph{IEEE Transactions on Visualization and
  Computer Graphics}, vol.~16, no.~5, pp. 729--741, 2010.

\bibitem[Cantlay(2007)]{cantlay2007gpu}
I.~Cantlay, \emph{{GPU Gems 3}}.\hskip 1em plus 0.5em minus 0.4em\relax
  Addison-Wesley Professional, 2007, ch. {High-Speed, Off-Screen Particles}.

\bibitem[Guennebaud et~al.(2007)Guennebaud, Barthe, and
  Paulin]{guennebaud2007high}
G.~Guennebaud, L.~Barthe, and M.~Paulin, ``{High-Quality Adaptive Soft Shadow
  Mapping},'' in \emph{Computer graphics forum}, vol.~26, no.~3.\hskip 1em plus
  0.5em minus 0.4em\relax Wiley Online Library, 2007, pp. 525--533.

\bibitem[He et~al.(2014)He, Gu, and Fatahalian]{He:2014}
Y.~He, Y.~Gu, and K.~Fatahalian, ``Extending the graphics pipeline with
  adaptive, multi-rate shading,'' \emph{ACM Trans. Graph.}, vol.~33, no.~4, pp.
  142:1--142:12, Jul. 2014.

\bibitem[nvi({\natexlab{a}})]{nvidiaMRS}
``{VRWorks -- Multi-Res Shading},''
  \url{https://developer.nvidia.com/vrworks/graphics/multiresshading},
  accessed: 2019-02-05.

\bibitem[nvi({\natexlab{b}})]{nvidiaLMS}
``{VRWorks -- Lens Matched Shading},''
  \url{https://developer.nvidia.com/vrworks/graphics/lensmatchedshading},
  accessed: 2019-02-05.

\bibitem[nvi({\natexlab{c}})]{nvidiaVRS}
``{VRWorks -- Variable Rate Shading (VRS)},''
  \url{https://developer.nvidia.com/vrworks/graphics/variablerateshading},
  accessed: 2019-02-05.

\bibitem[Vaidyanathan et~al.(2014)Vaidyanathan, Salvi, Toth, Foley,
  Akenine-M\"{o}ller, Nilsson, Munkberg, Hasselgren, Sugihara, Clarberg,
  Janczak, and Lefohn]{Vaidyanathan:2014}
K.~Vaidyanathan, M.~Salvi, R.~Toth, T.~Foley, T.~Akenine-M\"{o}ller,
  J.~Nilsson, J.~Munkberg, J.~Hasselgren, M.~Sugihara, P.~Clarberg, T.~Janczak,
  and A.~Lefohn, ``{Coarse Pixel Shading},'' in \emph{Proceedings of High
  Performance Graphics}, ser. HPG '14.\hskip 1em plus 0.5em minus 0.4em\relax
  Goslar Germany, Germany: Eurographics Association, 2014, pp. 9--18.

\bibitem[Vaidyanathan et~al.(2012)Vaidyanathan, Toth, Salvi, Boulos, and
  Lefohn]{Vaidyanathan:2012}
K.~Vaidyanathan, R.~Toth, M.~Salvi, S.~Boulos, and A.~Lefohn, ``{Adaptive Image
  Space Shading for Motion and Defocus Blur},'' in \emph{Proceedings of the
  Fourth ACM SIGGRAPH / Eurographics Conference on High-Performance Graphics},
  ser. EGGH-HPG'12.\hskip 1em plus 0.5em minus 0.4em\relax Goslar Germany,
  Germany: Eurographics Association, 2012, pp. 13--21.

\bibitem[Guenter et~al.(2012)Guenter, Finch, Drucker, Tan, and
  Snyder]{Guenter:2012}
B.~Guenter, M.~Finch, S.~Drucker, D.~Tan, and J.~Snyder, ``{Foveated 3D
  Graphics},'' \emph{ACM TOG}, vol.~31, no.~6, pp. 164:1--164:10, Nov. 2012.

\bibitem[Mittring(2007)]{mittring2007finding}
M.~Mittring, ``{Finding Next Gen: Cryengine 2},'' in \emph{ACM SIGGRAPH 2007
  courses}.\hskip 1em plus 0.5em minus 0.4em\relax ACM, 2007, pp. 97--121.

\bibitem[Bavoil et~al.(2008)Bavoil, Sainz, and Dimitrov]{bavoil2008image}
L.~Bavoil, M.~Sainz, and R.~Dimitrov, ``{Image-Space Horizon-Based Ambient
  Occlusion},'' in \emph{ACM SIGGRAPH 2008 talks}.\hskip 1em plus 0.5em minus
  0.4em\relax ACM, 2008, p.~22.

\bibitem[Williams(1978)]{williams1978casting}
L.~Williams, ``{Casting Curved Shadows on Curved Surfaces},'' in \emph{ACM
  Siggraph Computer Graphics}, vol.~12, no.~3.\hskip 1em plus 0.5em minus
  0.4em\relax ACM, 1978, pp. 270--274.

\bibitem[Reeves et~al.(1987)Reeves, Salesin, and Cook]{reeves1987rendering}
W.~T. Reeves, D.~H. Salesin, and R.~L. Cook, ``{Rendering Antialiased Shadows
  With Depth Maps},'' in \emph{ACM Siggraph Computer Graphics}, vol.~21,
  no.~4.\hskip 1em plus 0.5em minus 0.4em\relax ACM, 1987, pp. 283--291.

\bibitem[Fernando(2005)]{fernando2005percentage}
R.~Fernando, ``{Percentage-Closer Soft Shadows},'' in \emph{ACM SIGGRAPH 2005
  Sketches}.\hskip 1em plus 0.5em minus 0.4em\relax ACM, 2005, p.~35.

\bibitem[Ritschel et~al.(2009)Ritschel, Grosch, and
  Seidel]{ritschel2009approximating}
T.~Ritschel, T.~Grosch, and H.-P. Seidel, ``{Approximating Dynamic Global
  Illumination in Image Space},'' in \emph{Proceedings of the 2009 symposium on
  Interactive 3D graphics and games}.\hskip 1em plus 0.5em minus 0.4em\relax
  ACM, 2009, pp. 75--82.

\bibitem[Tomasi and Manduchi(1998)]{tomasi1998bilateral}
C.~Tomasi and R.~Manduchi, ``{Bilateral Filtering for Gray and Color Images},''
  in \emph{Sixth International Conference on Computer Vision, 1998}.\hskip 1em
  plus 0.5em minus 0.4em\relax IEEE, 1998, pp. 839--846.

\bibitem[McGuire(2017)]{McGuire2017Data}
M.~McGuire, ``{Computer Graphics Archive},''
  \url{https://casual-effects.com/data}, July 2017.

\end{thebibliography}

\end{document}